\documentclass[aps,prd,onecolumn,12pt,preprintnumbers]{revtex4-1}

\usepackage{hyperref}

\usepackage{float}

\usepackage{tikz}

\usepackage{graphicx}
\usepackage{epstopdf}

\begin{document}

\preprint{HDP: 16 -- 02}

\title{Physics of the Bacon Internal Resonator Banjo}

\author{David Politzer}

%\email[]{politzer@theory.caltech.edu}
\email[]{politzer@theory.caltech.edu}

\homepage[]{http://www.its.caltech.edu/~politzer}

%\thanks{452-48 Caltech, Pasadena CA 91125}
\altaffiliation{\footnotesize Pasadena CA 91125}
%\altaffiliation{\newline \em \em \em 452-48 Caltech, Pasadena CA 91125}
\affiliation{}

%\date{\today}
\date{June 17, 2016}

\begin{figure}[h!]
\includegraphics[width=4.0in]{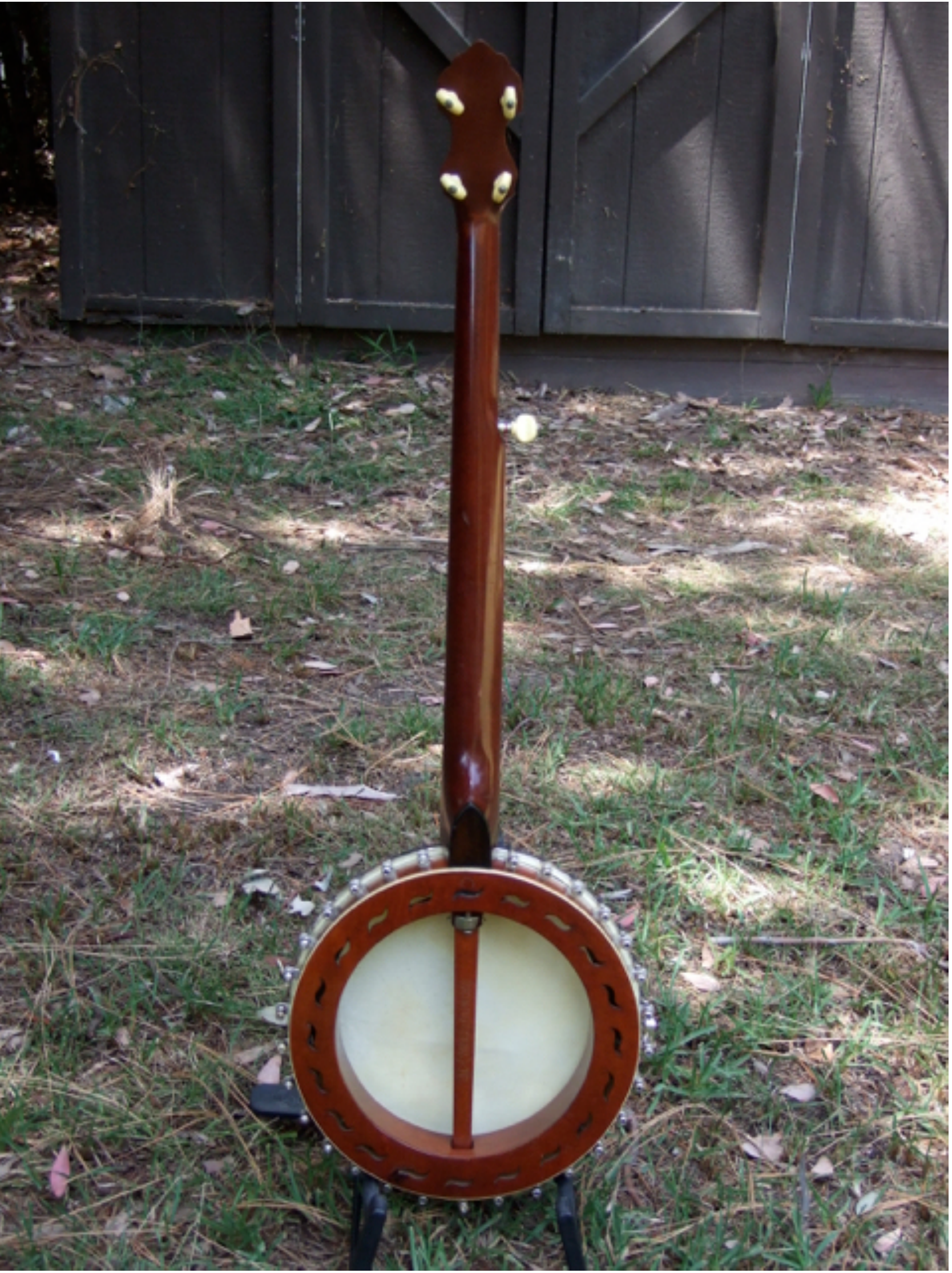}
\end{figure}

\begin{abstract}
The internal resonator banjo, patented and first sold by Fred Bacon around 1906, remains something of a cult favorite and is still produced by some independent luthiers.  According to enthusiasts, the characteristic design elements produce a sound that is mellower, richer, and of greater complexity and presence than without them.   Aspects of that sound are studied here, comparing instruments that are otherwise identical and identifying physics mechanisms that are likely responsible.

\end{abstract}

\maketitle{ {\centerline{\large \bf  Physics of the Bacon Internal Resonator Banjo}}

\section{Introduction}

In 1906 Fred Bacon, a virtuoso stage performer, patented\cite{patent} and then formed a company to produce and market what has come to be known as the internal resonator banjo.  In his patent application he wrote:

{\medskip  
\it ``This invention has relation to certain improvements in the construction of banjos or other similar musical instruments whereby a more lasting tone is produced and the quality of same improved.  The principal objection to the banjo resides in the fact that the tones are of short duration and that they therefore have a sharp staccato quality which is objectionable.  The object of this invention is to overcome this objection by providing the rim with a peculiarly-constructed annular chamber within which the partly-confined air can vibrate in harmony with the strings and cooperate therewith to produce a strong and resonant tone."}

\medskip
\noindent 
The design has its enthusiasts to this day and is still produced by independent luthiers.
\begin{figure}[h!]
\includegraphics[width=6.5in]{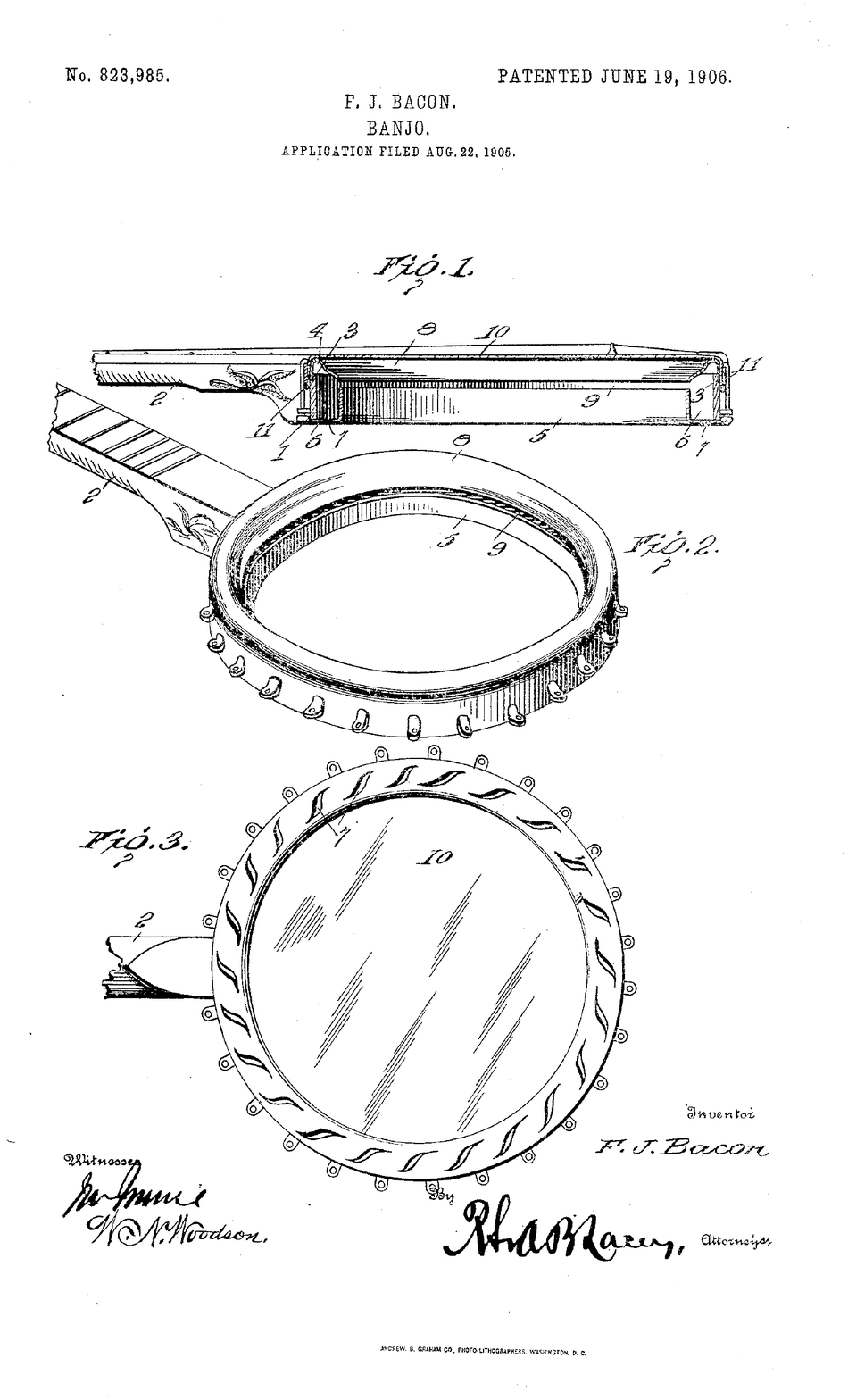}
\caption{Page 1 of Bacon's patent: \href{http://www.google.com/patents/US823985}{http://www.google.com/patents/US823985}}
\end{figure}

%\href{http://www.google.com/patents/US823985}{Bacon's patent}\cite{patent}

When I was offered my first real job, I went shopping for my first real banjo.  (That was 1977. I had been playing a banjo I'd built from scratch when I was 16.)  A local music shop carried some of the wonderful open-backs made by Kate Smith and Mark Surgies.  (They were the A.~E.~Smith Banjo Company).  However, my first real paycheck was months away, and I faced the expenses of moving across country and resettling.  Their top-of-the-line, modeled on the Bacon Professional ff, seemed an extravagance.  I settled for their open-back with a Bacon tone ring --- a fine banjo by any measure.  But I've been fascinated by internal resonators ever since.

To further my education in acoustics, a renowned expert strongly recommended Rayleigh's book on sound.\cite{rayleigh}  At one point, I came across his illustration, fig.~60, \S310, reproduced here as FIG.~2.   Rayleigh imagined that a double Helmholtz resonator might be somewhere of interest.  I imagined that it might be the key to the internal resonator.  I further realized that the annular chamber might also support certain frequencies of sound waves, not present in its absence, just as Bacon said.  It deserved closer scrutiny.

\begin{figure}[h!]
\includegraphics[width=4.5in]{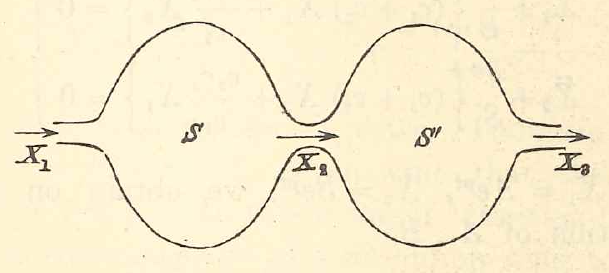}
\caption{from Rayleigh's {\it Theory of Sound}, fig.~60, \S310}
\end{figure}

\begin{figure}[h!]
\includegraphics[width=4.5in]{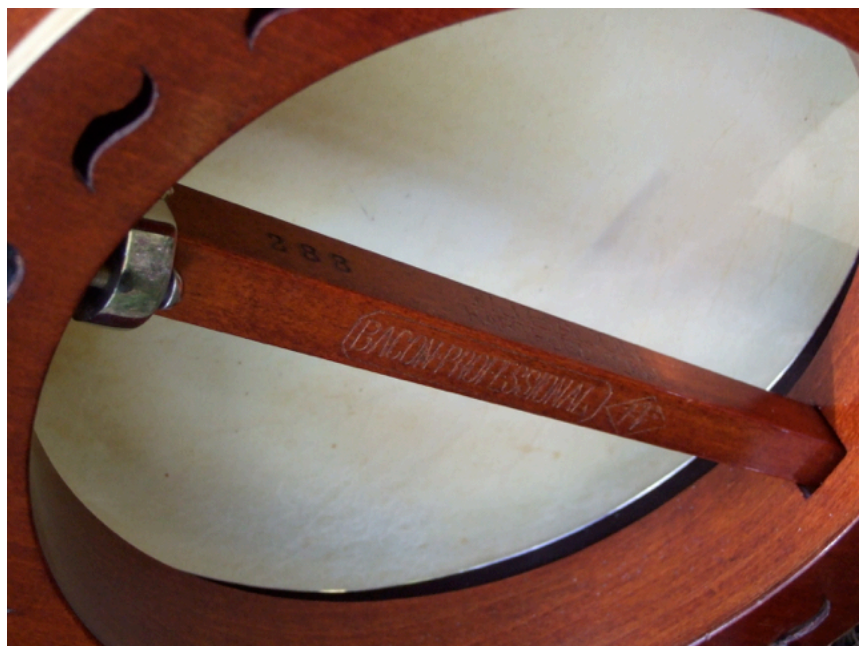}
\caption{The Bacon Professional ff internal resonator}
\end{figure}

%\newpage

\section{The basic physics ideas}

\begin{figure}[h!]
\includegraphics[width=6.5in]{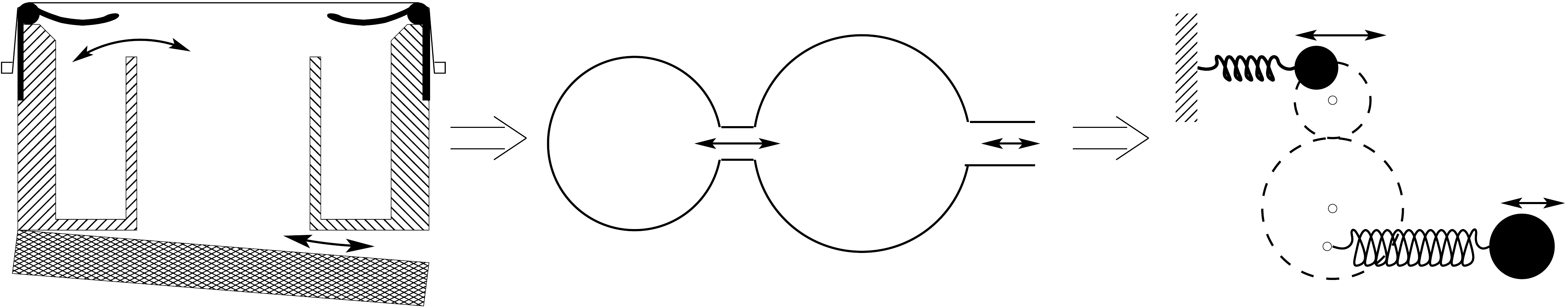}
\caption{The double-Helmholtz interpretation of the internal resonator}
\end{figure}

A partial back and a cylindrical wall, extending from the back to close to but not touching the head, divide the cylindrical volume of the pot (the banjo body) into an inner cylinder and a surrounding annular region.  The left-hand drawing in FIG.~4 is a schematic representation of a cross section of the pot.  The two volumes are connected by a constricted region formed by the top of the wall and the tone ring.  And the exit to the outside air is a second constriction formed by the partial bottom and the belly of the player.  The central drawing of FIG.~4 is an idealization in terms of coupled Helmholtz resonators.  And the right-hand figure is the mechanical analog:  The black circles are the masses of air in the constricted regions, and the springs represent the compressible air in the large volumes.  The two dashed circles are meant to represent ideal gears.  They reflect the fact that the two constricted regions need not have the same area as they open into the central cylinder.  Hence, it has the potential to act something like a hydraulic lift.  Were this picture applicable, there would be two Helmholtz resonances, whose frequencies could be selected over a very wide range by adjusting the constriction dimensions.  However, careful measurements show that the actual internal constriction is necessarily too large for this interpretation, and there is really only one Helmholtz resonance.  In agreement with the basic features of Helmholtz resonators, its frequency is distinctly lower than that of a simple rim because of the partial back.

Further experiments confirm that Bacon was right about the higher frequency resonances of the air in the pot.  The combined system can be well-represented as a coupling of two well-understood systems: a smaller central cylinder and an annulus of rectangular cross section.  While the problem of waves in an annulus is not exactly soluble, it is certainly close to a rectangular cross-section pipe with identified ends\cite{periodic-boundary} --- at least for the dimensions that appear in the internal resonator banjo.  This coupled system has a richer spectrum of resonances, starting at a lower frequency than what is available with the simple rim.

This report gives the details of these results and describes the observations that support them.  Head motion is the primary producer of banjo sound.  In addition to the force of the strings {\it via} the bridge, air motion inside the pot produces pressure variations that push on the head.  So, banjo timbre is subtly effected by pot air dynamics.

\section{alternate physics strategies}

For many purposes, faithful mathematical modeling is of enormous value.  In such cases, all available resources are brought to bear on the problem so that the math description is as detailed and accurate as possible.  On the other hand, having a simple way to think about a system has its virtues.  For some people, it's just very satisfying.  But, even on the most practical level, simple but valid pictures can help greatly with creating novel designs and finding solutions to particular practical problems.  This project is very much along the lines of the simple pictures approach.

\section{Outline} 

As already alluded to in the discussion pertaining to FIG.~4, several lines of inquiry did not pan out.  The order of presentation herein reflects what I think to be the simplest explanation of the conclusions I finally drew.  It is not the actual order in which the study unfolded.

Section V is an overview of Helmholtz resonators, and section VI is a reminder and summary of an accompanying, separate study of Bacon's tone ring on its own.  Section VII presents the evidence that there is only one Helmholtz resonance (and not two) for the internal resonator construction.    Section VIII studies the air vibrations  where air sloshes from place to place within the pot, identifying the particular contribution of the internal resonator annular region.  Fully assembled banjos are compared in section IX, including plucks, spectra, spectrographs, and frailed sound samples.  And section X concludes with the lessons learned.

\section{Helmholtz Resonators}

Understood broadly, the concept of the Helmholtz resonance is of enormous value in many acoustical settings.  There is, indeed, a simple formula that applies to a very idealized situation.  But interpreted qualitatively, that model suggests a way to view and understand many different, important systems.  Helmholtz ``invented" them to serve as very sensitive, narrow-band detectors of sound.  However, they have been used to absorb sound in architectural and engineering settings and to produce sound in musical settings since time immemorial.  For most stringed instruments,  they enhance the sound of the lowest frequencies.  The common feature is that a relatively small volume of air is pushed back and forth by the expansion and contraction of a much larger volume.

\begin{figure}[h!]
\begin{center}
\begin{tabular}{ cc }
\includegraphics[width=2.5in]{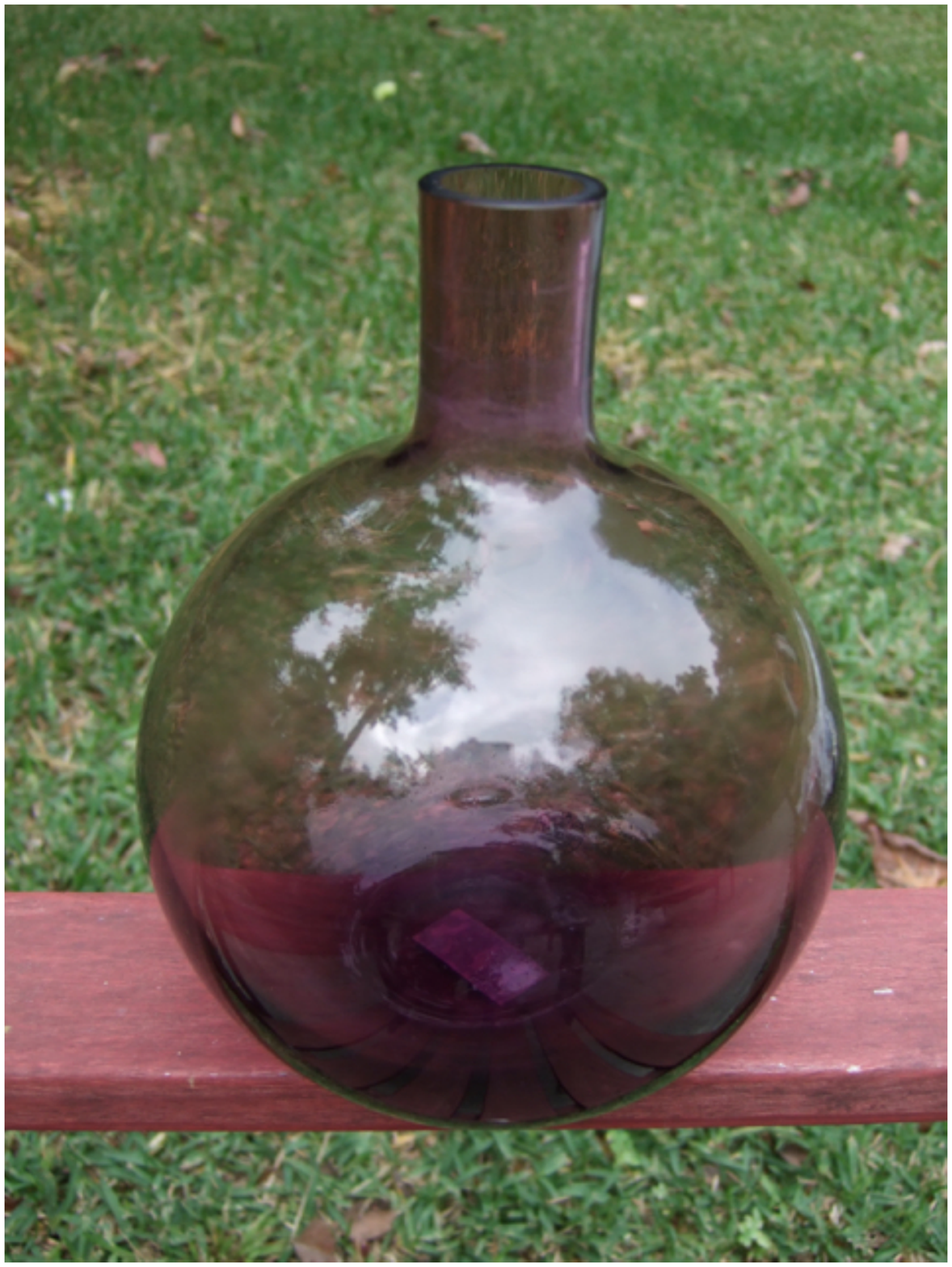}
&
\includegraphics[width=3.0in]{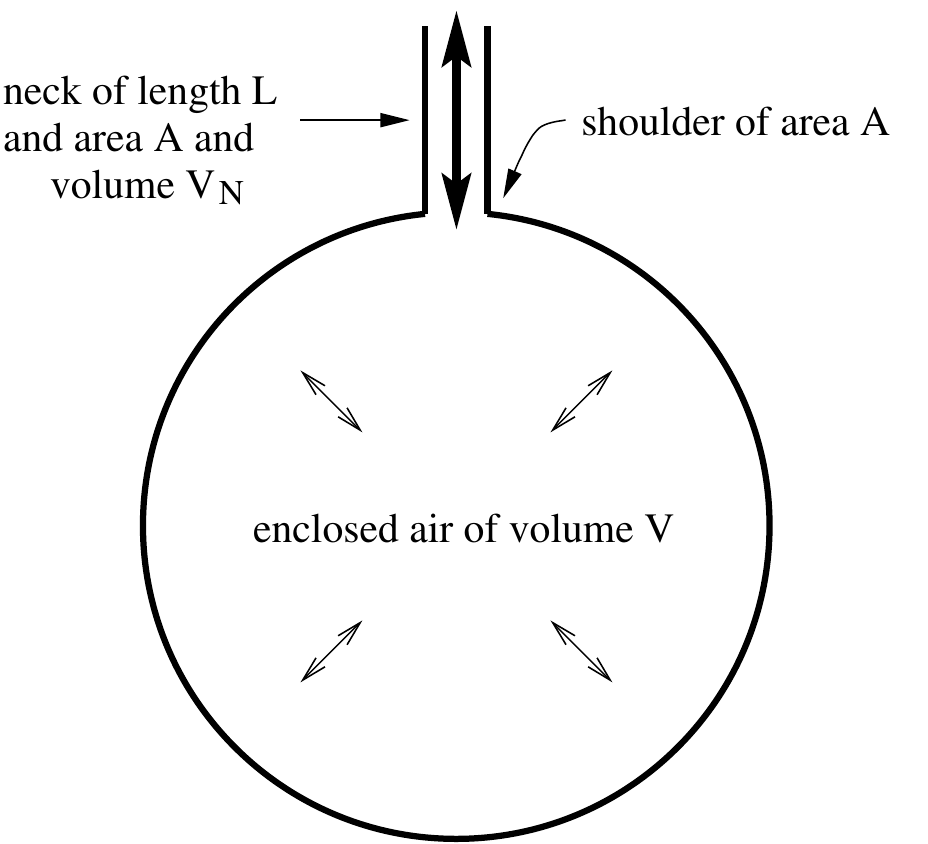}
\end{tabular}
\end{center}
\caption{Helmholtz bottle resonators: real \& ideal}
\end{figure}

In the idealized version, the volume, $V_N$, of air in the neck of the imagined bottle (see FIG.~5) determines the mass of the oscillator.  The larger volume, $V$, produces the springiness.  The Hooke's Law ``spring constant" is actually proportional to $A^2 / V$, where $A$ is the area of the interface.  (If the bottle neck is cylindrical, then $V_N = A \times L$, where $L$ is the length of the neck.)  The resonant frequency, $f_H$, is given by

\medskip

\centerline{$f_H = ${\Large {${v_s \over {2 \pi}}  \sqrt{A^2 \over {V \times V_N}}$}}}

\medskip

\noindent
where $v_s$ is the speed of sound --- which characterizes the inherent springiness of air.  To the extent that this idealization is applicable, these are the only parameters that enter into the determination of $f_H$.  In particular, variations in shape and positioning have no appreciable effect.

The qualitative features of this formula account for the behavior of any situation where a relatively large, enclosed volume opens onto the open air through some sort of constriction.  But even the real bottle shown in FIG.~5 presents problems with taking the numbers too seriously.  Its volumes can be measured with water and a measuring cup.  The diameter of the cylindrical neck can be measured with a ruler.  But how long is the neck, and what is the value of area $A$?  For a long pipe, it is known and qualitatively understood that the pipe length effectively extends into the open air by 0.3 to 0.4 times the pipe diameter.  That ``extra" length represents the mass of outside air that participates significantly in the back-and-forth motion. If the ``neck" is, in fact, a stringed instrument sound hole, even a rough estimate of the effective $V_N$ presents a significant challenge.  And what about the bottom of the neck?  The real bottle in FIG.~5 has a smooth, curving transition.  That introduces a small ambiguity in $V_N$ but a rather large ambiguity in $A$.  The effective $A$ is clearly larger than the cross section of the strictly cylindrical upper portion of the neck --- but not by too much.

The pitch of the sound produced by blowing across the opening of the particular bottle in FIG.~5 was 122 Hz.  Using water, a measuring cup, and a ruler, adding the canonical open-end correction, and making an educated guess for $A$, I got 127 Hz from the formula.  Of course, there exists a value for $A$ for which the formula gives exactly the right answer, but, working {\it a priori}, one can only guess what that might be.

The bottle as a musical instrument is typically tuned by fine adjustment of $V$, e.g., by partially filling the bottle with water.\cite{danish}

The qualitative dependence of $f_H$ on $A$ is used in many musical instruments.  The ocarina and its cousins are examples.  (See FIG.~6.)  Their frequency spectra contain very little besides the basic Helmholtz resonance.  Enlarging the escape area raises the pitch. 

\begin{figure}[h!]
\begin{center}
\begin{tabular}{ cc }
\includegraphics[width=4.4in]{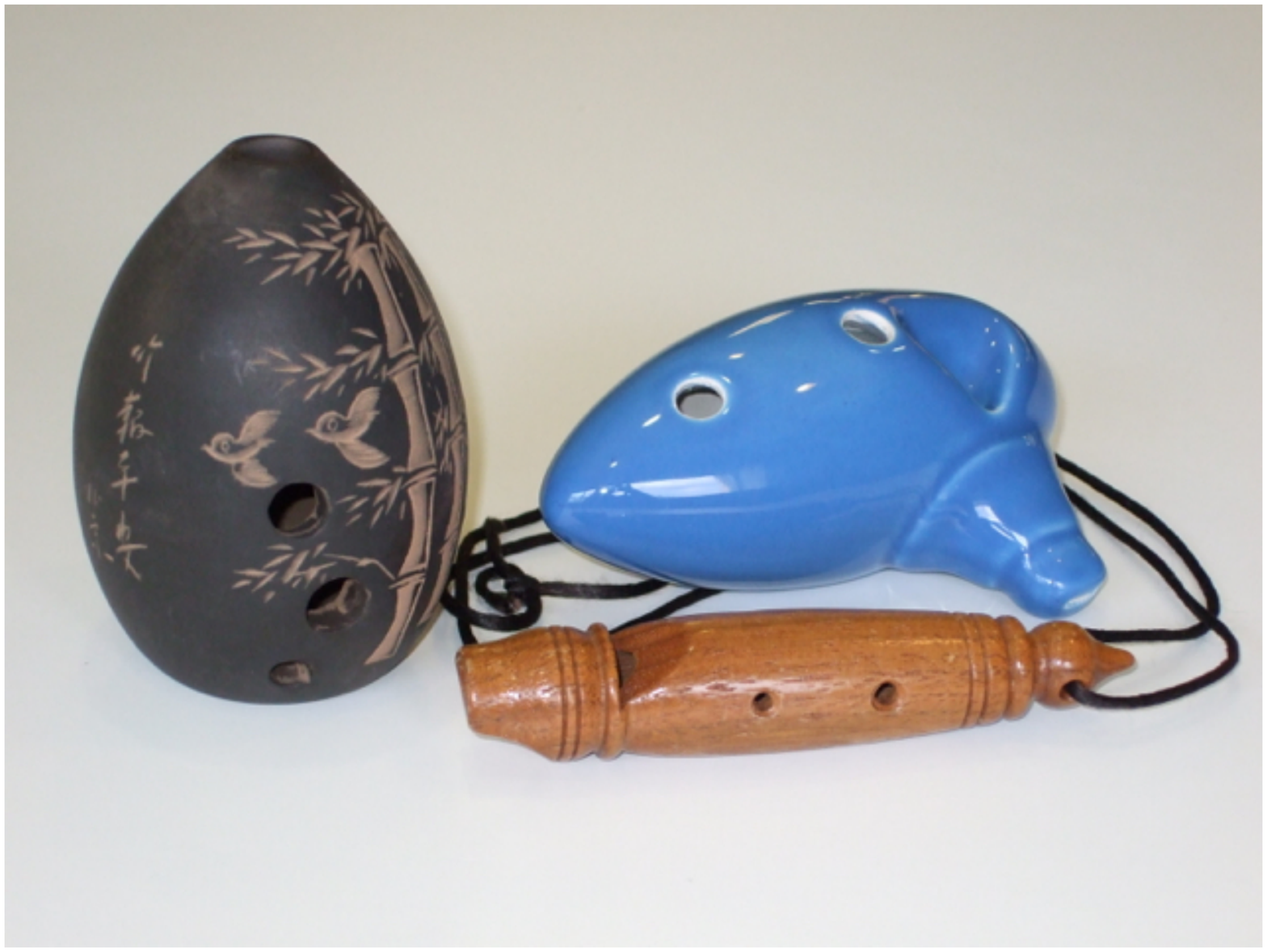}
&  
\includegraphics[width=2.35in]{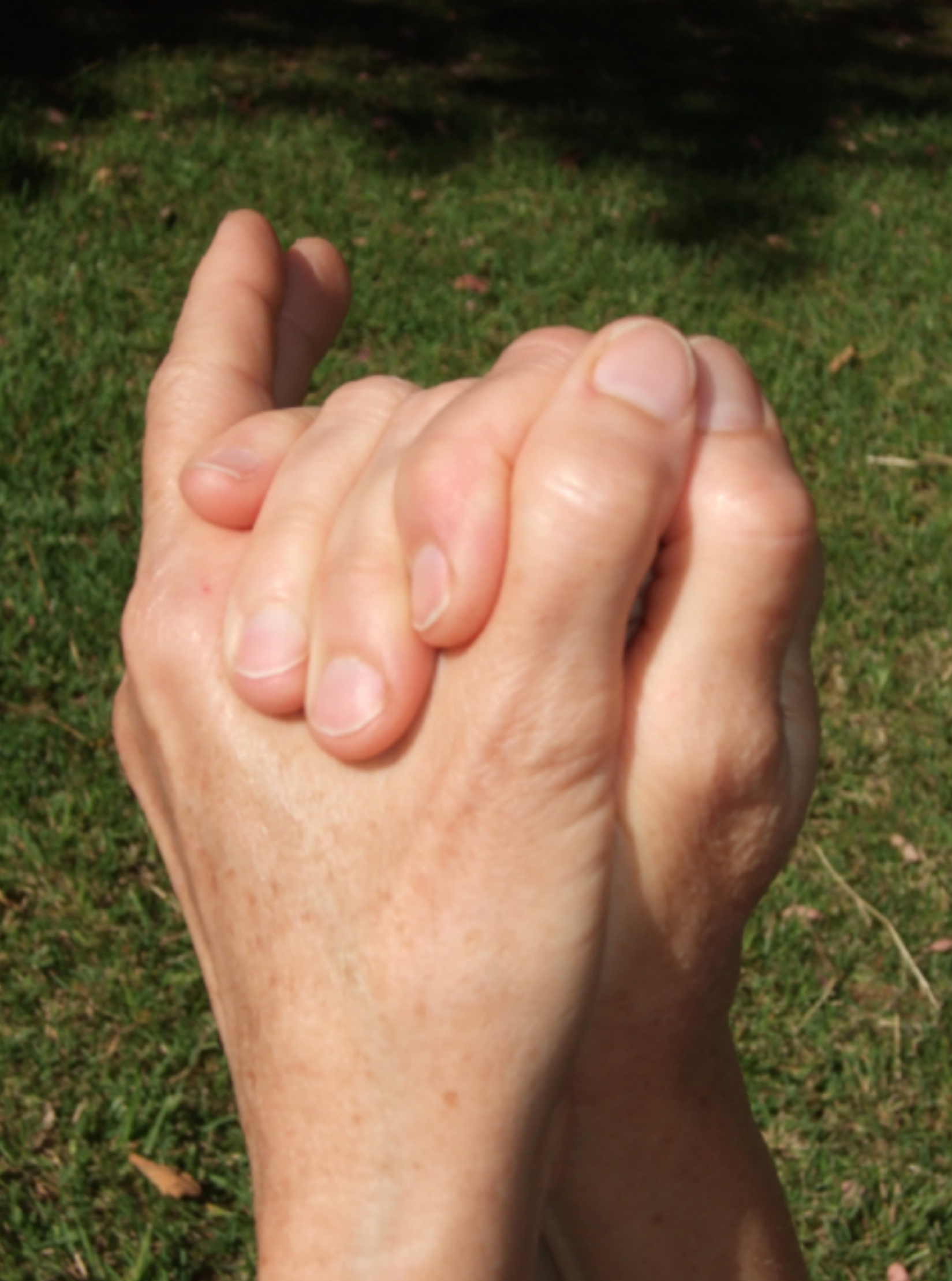}
\end{tabular}
\end{center}
\caption{ Ocarinas: Chinese xun, Bryan Mumford's Puny Tune, and Darryn Songbird's Sweet Potato (left) \& bare hands (right)}
\end{figure}

In the context of banjos, enlarging $V_N$ lowers the pitch.  In the design and set-up of standard resonator banjos, attention is paid to this detail.  In the context of open-back banjos, not only can $V_N$ be made bigger but $A$ can be made smaller.  Both are accomplished by a partial back.  FIG.~7 shows a 100 year-old six-string banjo with just such a back.  Note that the player's belly is a crucial part of the open-back banjo's Helmholtz resonator.\cite{openback}  $V$ is the volume inside the pot, $V_N$ is roughly the volume between the partial back and the player's belly, and $A$ is the region between the two $V$'s.  Lowering the standard open-back Helmholtz frequency was clearly the goal of the builder of this 100 year-old banjo to support the sound of the additional low string.

\begin{figure}[h!]
\includegraphics[width=5.5in]{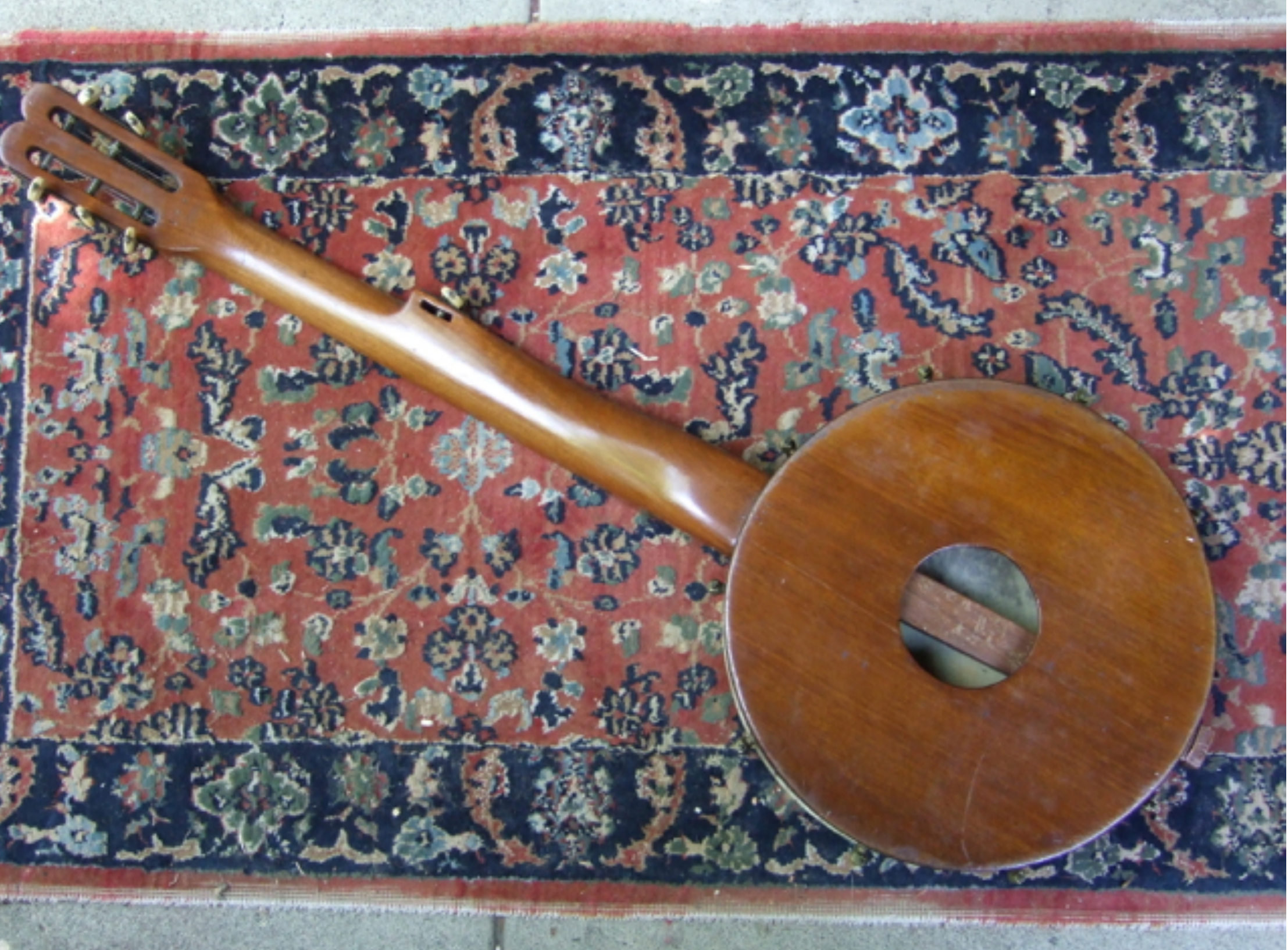}
\caption{A 100 year-old, flush-fret six-string banjo}
\end{figure}

For a banjo, sound production from the Helmholtz resonance is not just the in-and-out air.  It is also heard in its effect on motion of the head (which makes most of the  sound) as that motion is altered by the up-and-down pressure of the Helmholtz resonance from underneath.

\section{The Bacon Tone Ring}

Bacon included a metal tone ring in his design.  This item, shown in cross section in FIG.s~4 and 8, sits on the top edge of the wood rim, and the drum head is stretched over it.  A detailed study of its vibrations and their effect on the banjo's sound is presented in ref.~\cite{tone-ring}.  In summary: the $1/4''$ solid diameter core hardens the rim edge to reduce high frequency absorption relative to a pure wood rim.  The two thin flanges stiffen the rim (at the cost of very little extra weight) to reduce large rim motions that otherwise absorb low frequencies.  And vibrations of the free horizontal flange absorb some of the vibrational energy that without the flange would have gone into sound.

\begin{figure}[h!]
\includegraphics[width=2.0in]{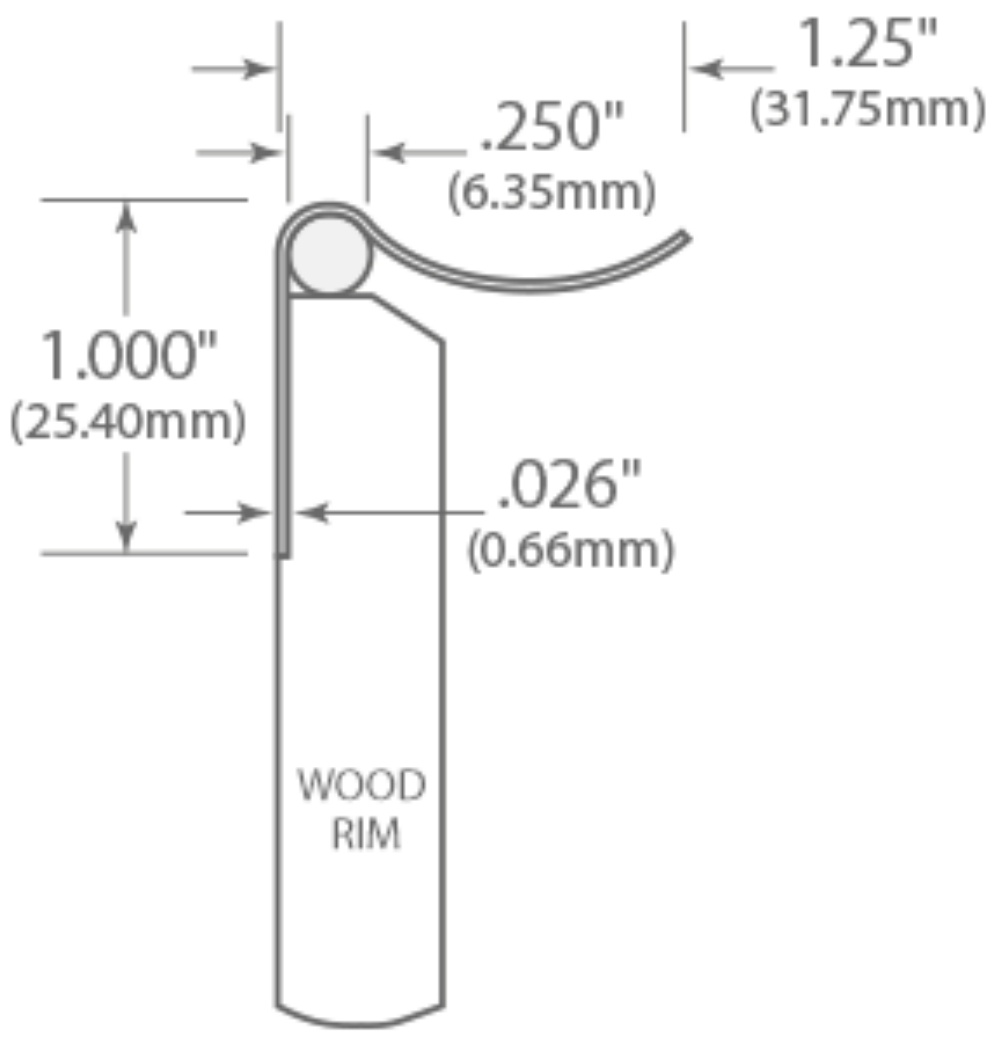}
\caption{The Steward-MacDonald reproduction Bacon tone ring, discussed in some detail in ref.~\cite{tone-ring}}
\end{figure}

In the present study, the focus is on air vibrational modes of the pot.  The tone ring is considered, at least theoretically, as simply a fixed (and ultimately irrelevant) part of the defining geometry.  Almost all of the comparisons are made with a Bacon tone ring installed on a single $11''$ Goodtime\cite{Goodtime} rim.  The pot geometry and dimensions are altered by attaching a variety of  backs and internal resonators.  Whatever stiffening and vibrating the tone ring does, it does so similarly for the different back and internal geometries.    Only the comparisons in section IX involve an all-wood, standard  Goodtime rim, where it is contrasted with a Bacon-like Goodtime, i.e., with tone ring and full-size internal resonator installed.  This comparison involves all the effects at once, and is, so to speak, the proof of the pudding.

\section{Internal Resonator Helmholtz Resonance}

To disentangle the various mechanisms that may be at play with an internal resonator, first focus on the Helmholtz resonance(s).  The situation is further simplified by beginning with partial backs with different size holes.  The bottom of a Bacon-tone-ring-equipped Goodtime rim was cut flat and fitted with six threaded inserts.  The rings shown in FIG.~9 were cut from 3.0 mm, 7-ply birch and could be attached with a narrow retaining ring of the same plywood and six screws.

\begin{figure}[h!]
\includegraphics[width=4.5in]{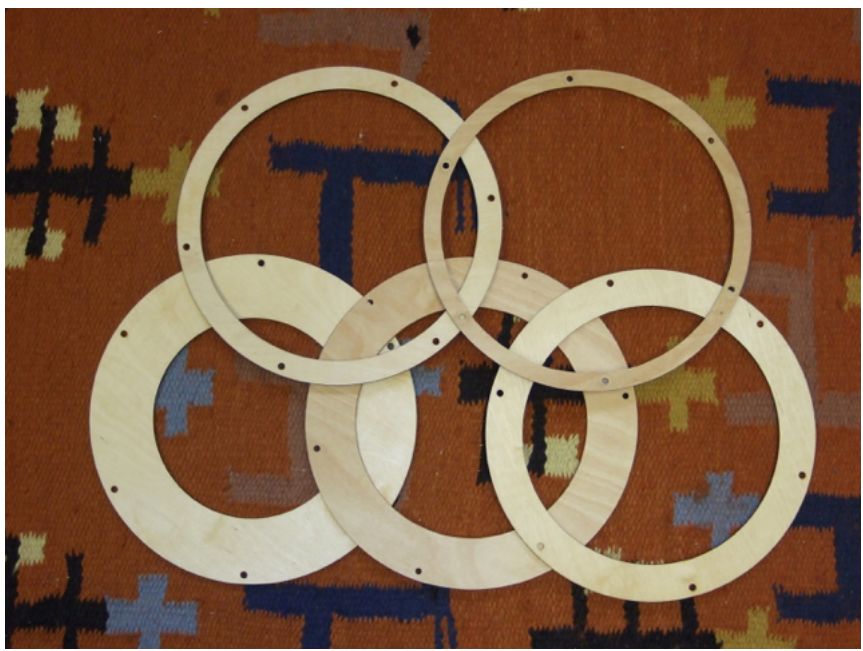}
\caption{Rings with various hole diameters that can be attached to the back}
\end{figure}

With strings, neck, and tailpiece removed (but coordinator rod in place), the sounds of  head taps with a piano hammer were recorded for various bottom hole diameters.   The largest was the stock Goodtime, whose inner diameter is 9 $3/4''$.  The smallest was 7 $5/8''$, which is the diameter of all the internal resonator inserts  that appear later.

I mounted a synthetic belly, made of closed cell foam, cork, and Hawaiian shirt.  (Those materials mimic the absorption and reflection of a player's body).    The opening to the outside air is chosen to approximate typical playing and is far more reproducible than holding the instrument up to one's body.  The genesis and details of this back are discussed in ref.~\cite{openback}.  Since a Helmholtz resonance is characterized by motion of air in and out of the sound hole, I placed the microphone right at the opening.  So the microphone placement helps focus on the specific modes of interest.  Most of the sound actually comes off the head and is predominantly due to other modes.

FIG.~10 shows the spectra for long series of those head taps, plotted for 100 to 500 Hz.  The vertical scale is decibels, a logarithmic measure of the sound pressure.  (Were the scale linear in pressure, the individual resonances would look more striking.)

\begin{figure}[h!]
\includegraphics[width=6.5in]{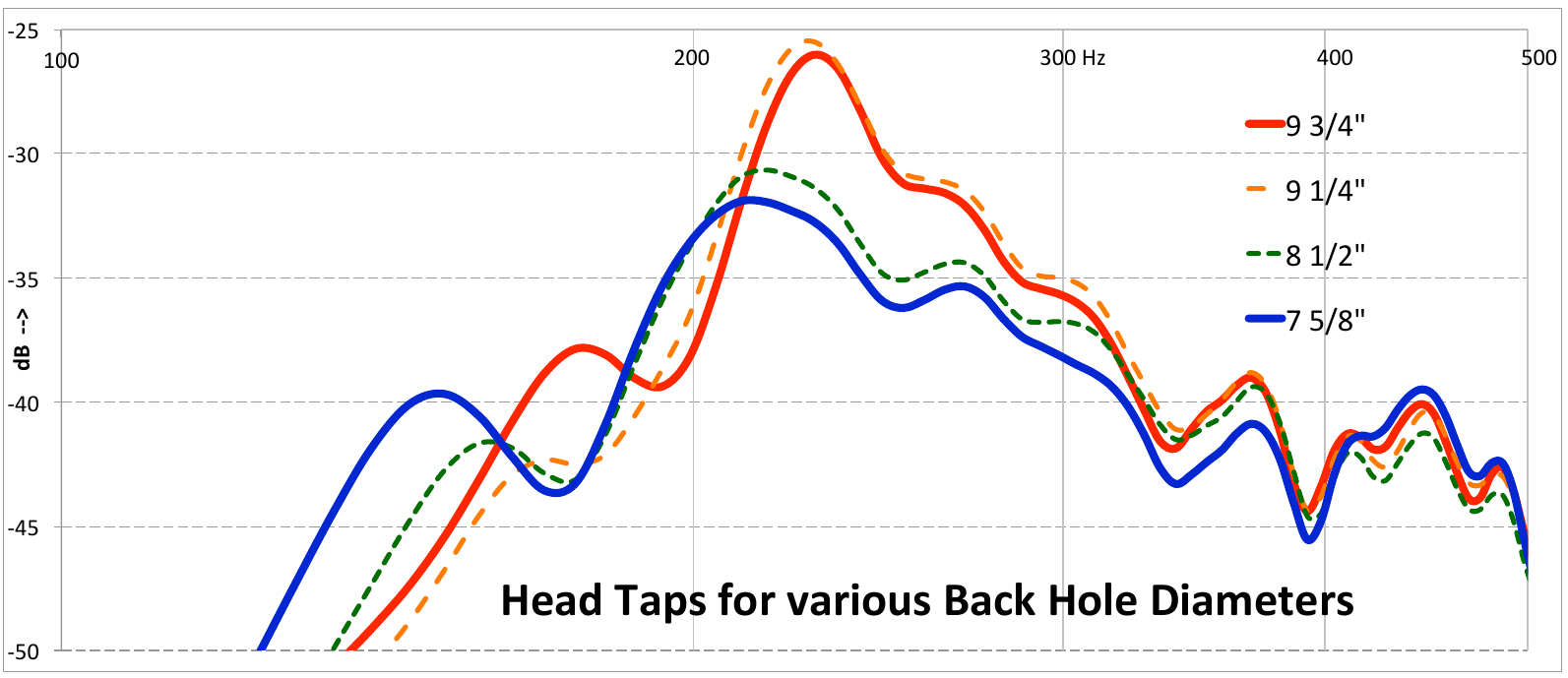}
\caption{head taps with foam belly-back; curves labeled by back hole diameter (no internal wall)}
\end{figure}

The strongest resonance for each back is the one with the second lowest frequency.  It and the next lower one show a systematic decrease in frequency with decreasing back hole diameter.  The higher  resonances (at least six of them) show no appreciable frequency dependence on the back hole dimension.

This is precisely the qualitative behavior to be expected from the Helmholtz formula.  Smaller hole diameter implies smaller $A$ and larger $V_N$ in the formula for $f_H$.  There are two peaks for each back that reflect this behavior because the internal pot Helmholtz resonance couples strongly to the lowest drum mode of the head.  Not only do they both push the same plug of air in and out, but they also push on each other over the whole head surface.  The higher frequency modes are due to other physics.  It is typical that the lowest two modes of the body of a stringed instrument are the coupled versions of the Helmholtz and lowest sound board modes.  And, on typical banjos, the fundamental frequencies of all strings but the high $5^{\text{th}}$ are below 300 Hz.  Also, on the banjo, the coupling of the two low modes is particularly strong because the head moves a lot compared to wood sound boards.

\begin{figure}[h!]
\includegraphics[width=4.5in]{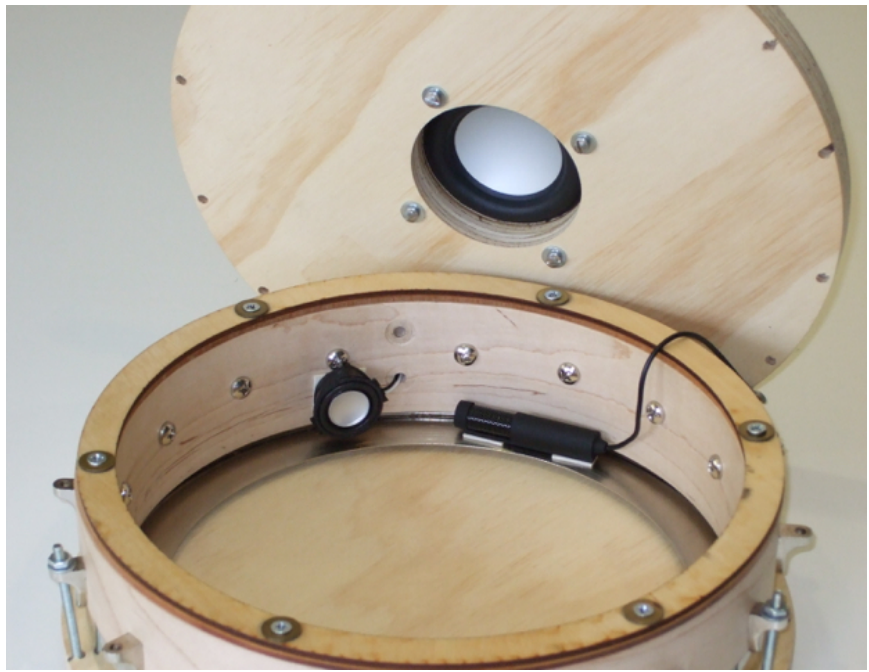}
\caption{plywood heads, speakers, \& mic}
\end{figure}

To separate Helmholtz from sound board physics on violins and guitars, experimenters have occasionally buried the instrument in sand -- to immobilize the sound board motion.  It's easier on the banjo.  I simply replaced the regular head with $3/4''$ plywood.  To drive the Helmholtz resonance, I mounted a $3''$ speaker in the middle of that plywood head.  That head is the one {\it not} attached to the rim in FIG.~11 and installed on the rim in FIG.~12.  (The attached solid head and rim-mounted $1''$ speaker and microphone in FIG.~11 are described in section VIII.)  The speaker is driven with a signal generator and audio amplifier with a slow sweep, logarithmic in frequency, over the desired ranges.

\begin{figure}[h!]
\includegraphics[width=3.5in]{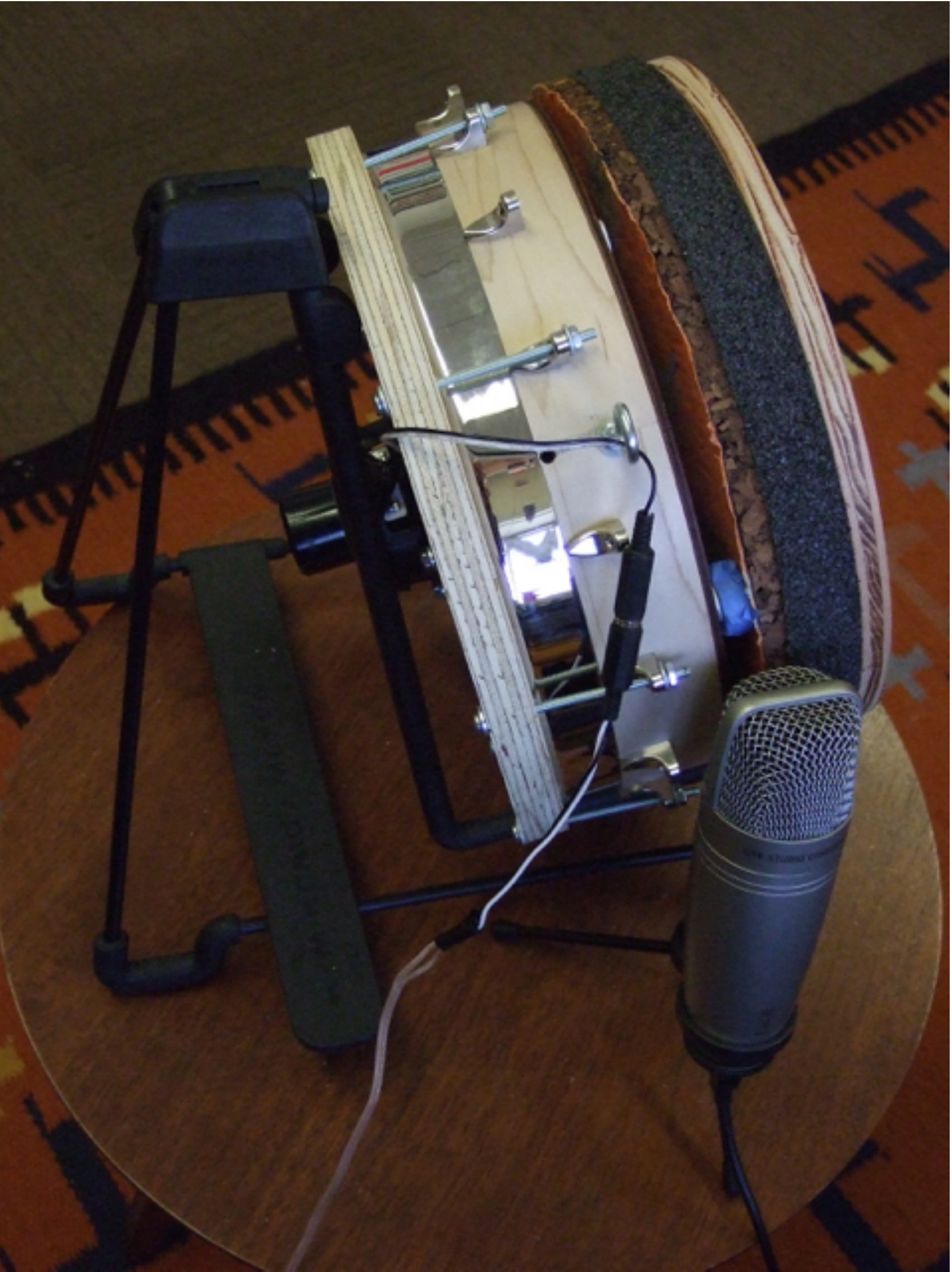}
\caption{wood ``head" with $3''$ speaker and cork \& foam belly-back}
\end{figure}

\begin{figure}[h!]
\includegraphics[width=5.7in]{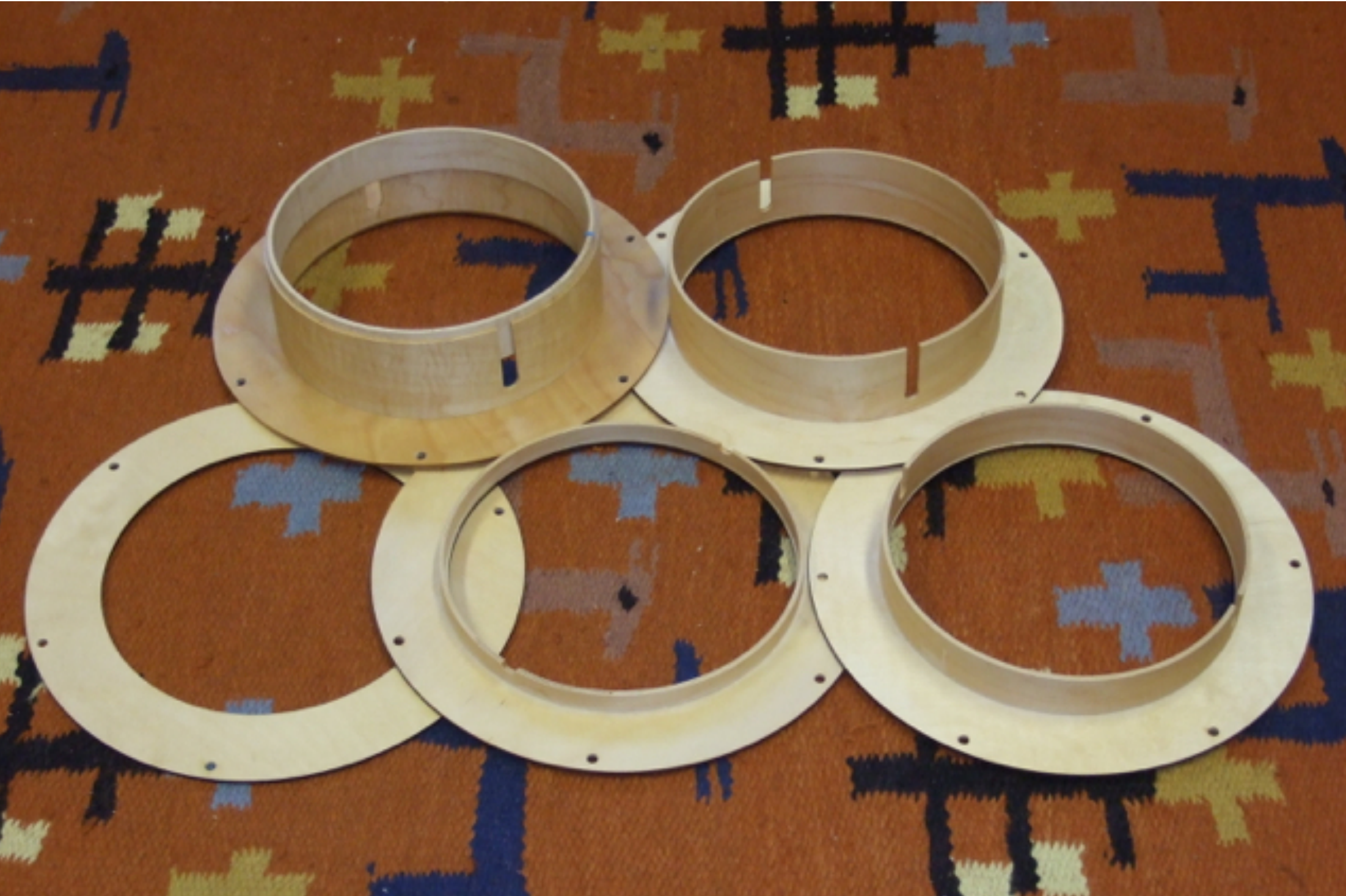}
\caption{internal resonators of various and variable heights}
\end{figure}

So, the frequencies of the lowest two pot resonances are substantially lowered by the partial back in a way whose physics is qualitatively understood.  The next question is the impact of the cylindrical wall of the internal resonator that divides the interior into a smaller, central cylinder and an outer annular volume.  I fabricated a variety of internal resonators, shown in FIG.~13, all with the same cylinder diameter and back hole size but with various heights.  The cylinders were cut from 3.0 mm, 5-ply maple drum shell stock.

It turned out that what was really needed to reduce the physics to something simple and obvious were internal walls that came yet closer to the inner surface of the plywood head.  Rather than fabricating a new set, I made a split ring that could be inserted into the highest original cylinder.  It could be placed carefully at any particular distance from the head when assembled and tightened snugly with a shim in the gap in its circumference.  That is the upper left construction in FIG.~13.

The resulting spectra for driving with the $3''$ head-mounted speaker and listening with a microphone at the rim-belly-back opening are plotted in FIG.~14.  Now, for each pot geometry, there is only one, low, broad peak between 200 and 300 Hz.  With this set-up, the higher resonances are all considerably weaker.  All versions are with the same rim with its Bacon tone ring.  The black curve, labeled ``stock" refers to the standard, open-back Goodtime rim.  The red curve, labeled ``7 $5/8''$" ring, is the partial back with no cylindrical wall.  (That ``7 $5/8''$" is the same partial back hole size as all of the internal resonators.)  The blue and green curves refer to internal resonators that have a $3/8''$ and  $1/4''$ space, respectively, between the top of the internal cylinder and the inner surface of the head.  The pale blue ``no gap" curve refers to an inner cylinder that touches the head and seals off the outer annulus from the inner cylinder.

\begin{figure}[h!]
\includegraphics[width=6.0in]{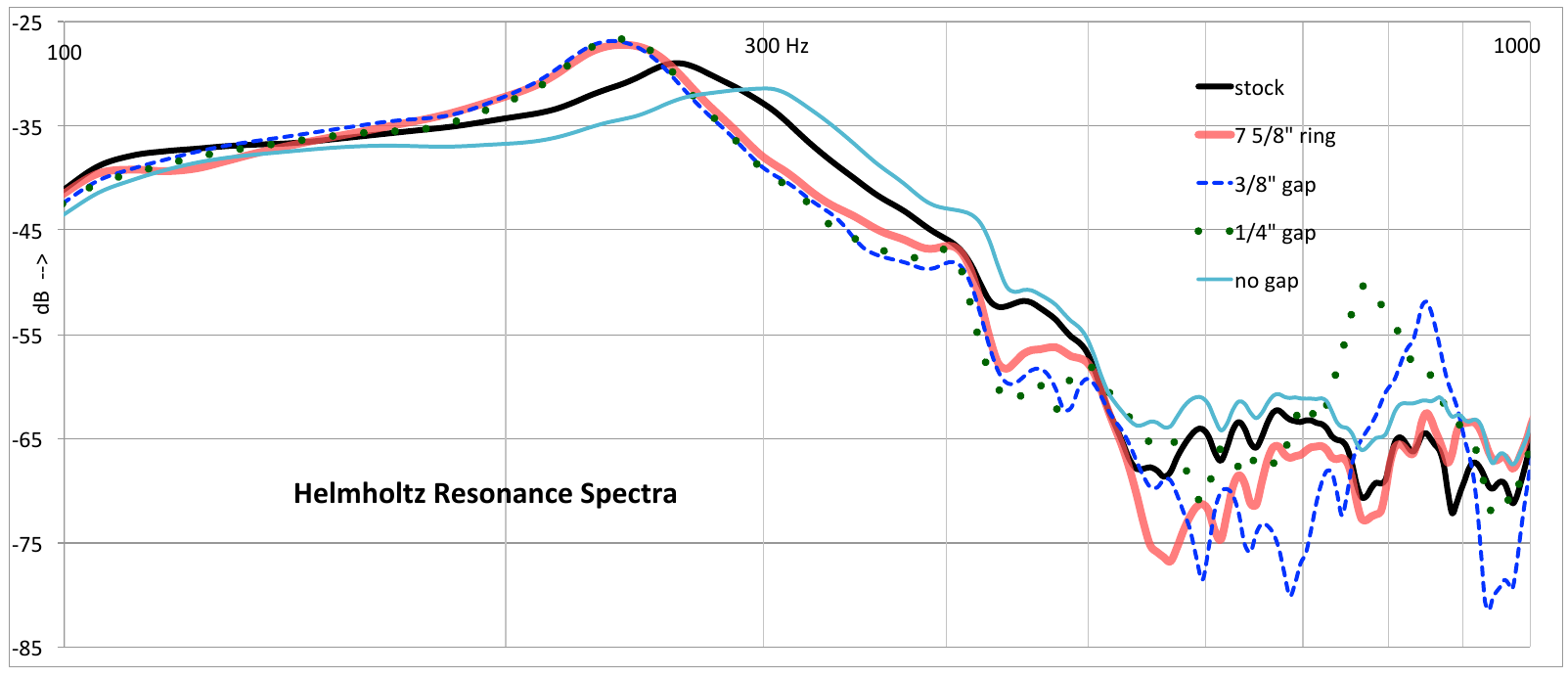}
\caption{%Helmholtz-spectra-100-1kHz-8192.xlxs; 
plywood head with $3''$ speaker; foam belly-back}
\end{figure}

The relations between the red, black, and pale blue curves are standard Helmholtz resonator physics.  The black (stock) and red (7 $5/8''$ ring) have that same $V$, but black has a smaller $A$ and a larger $V_N$.  The black and pale blue (no gap) have the same $A$ and $V_N$, but pale blue has a smaller $V$.  Black and pale blue differ in all three parameters; so the sign of the difference depends on details of the actual values.

A very important lesson from these measurements, which was not altogether obvious beforehand, is that the Helmholtz resonances of the red, blue, and green curves, i.e., all of the pots with the same size partial back, are essentially indistinguishable.  That means that the height of the internal cylindrical wall, going from zero all the way up to the height in the standard, finished banjo (i.e., reaching to $3/8''$ from the inner surface of the head) does not effect the frequency of the Helmholtz resonance.  They are all the same --- as if there were no inner wall at all.   On one hand, the simple Helmholtz resonance picture says that the resonant frequency is independent of the shape of the cavity.  So, apparently, the wall is simply an alteration in the shape.  And these three configurations have the same $V$, $V_N$, and $A$.  On the other hand, one might ask whether there could be a wall sufficiently high that it divides the original cavity into two serial Helmholtz resonators --- just as Rayleigh suggested could arise, at least for some design.  Apparently, in practice, the answer is no, not for the internal resonator geometry.  There are two obstacles.  Friction becomes an important force with much smaller gaps.  And the volume of the purported neck is too small relative to the interface area.

\section{Cavity Modes}

The internal wall certainly {\it does} something, and that is revealed by a study of the higher frequency cavity modes.  Again, the coupling to the head modes is removed by using a solid $3/4''$ plywood head.  That is the solid disk in FIG.~11.  Since these air modes are essentially internal to the pot, the ``sound hole" gap can be eliminated --- to allow for cleaner and clearer resonances.  The sound hole was only crucial to the Helmholtz mode.  So I chose to seal the back with solid plywood.  And that required putting a driving speaker and a recording microphone inside the pot.  That assembly is shown in FIG.~11.

\begin{figure}[h!]
\includegraphics[width=6.5in]{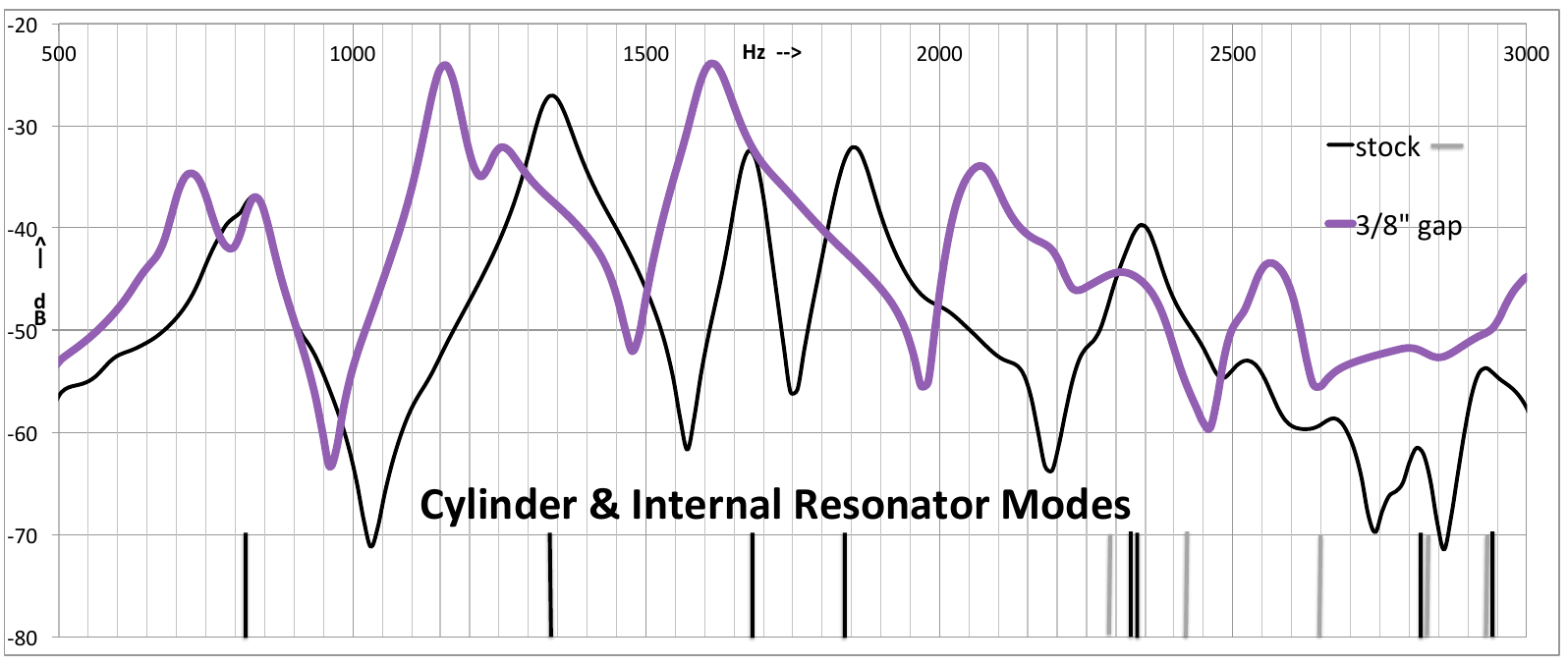}
\caption{Comparison of the stock pot \& the internal resonator with a $3/8''$ head-to-resonator-wall gap; lines at the bottom denote frequencies calculated solely from the actual pot dimensions}
\end{figure}

Again, the rim is the Goodtime fit with a Bacon tone ring.  Slow, logarithmic frequency scans, much as before but now with the small, internally mounted speaker and microphone, yielded the spectra shown in FIG.~15.  Note that the horizontal frequency scale is linear.  ``Stock" refers to the standard rim. ``$3/8''$ gap" is the standard internal resonator, whose cylindrical wall is 2 $1/4''$ high, which brings it to $3/8''$ from the inner surface of the head.  The are no Helmholtz resonances in this configuration because there is no in-and-out air motion.  The lowest closed cavity resonances are the ones shown.

At least below 1800 Hz, the internal resonator resonances appear to have been shifted  lower, and there are more of them.  The challenge is to understand their origin.

The first step is to recognize that the stock pot resonances are standard fare in undergraduate physics and even are often used as a pedagogical experiment in laboratory courses.  For an ideal cylinder the modes and frequencies can be calculated. They are presented in many places.\cite{cylinder-modes}  Pressure node lines and frequencies are shown in FIG.~16 for the ``squat" cylinder.  ``Squat" means that the dimension perpendicular to the circular cross section is sufficiently small that there is no pressure variation in that perpendicular direction until yet higher frequencies where half a wave fits between the head and the back.

\begin{figure}[h!]
\includegraphics[width=5.5in]{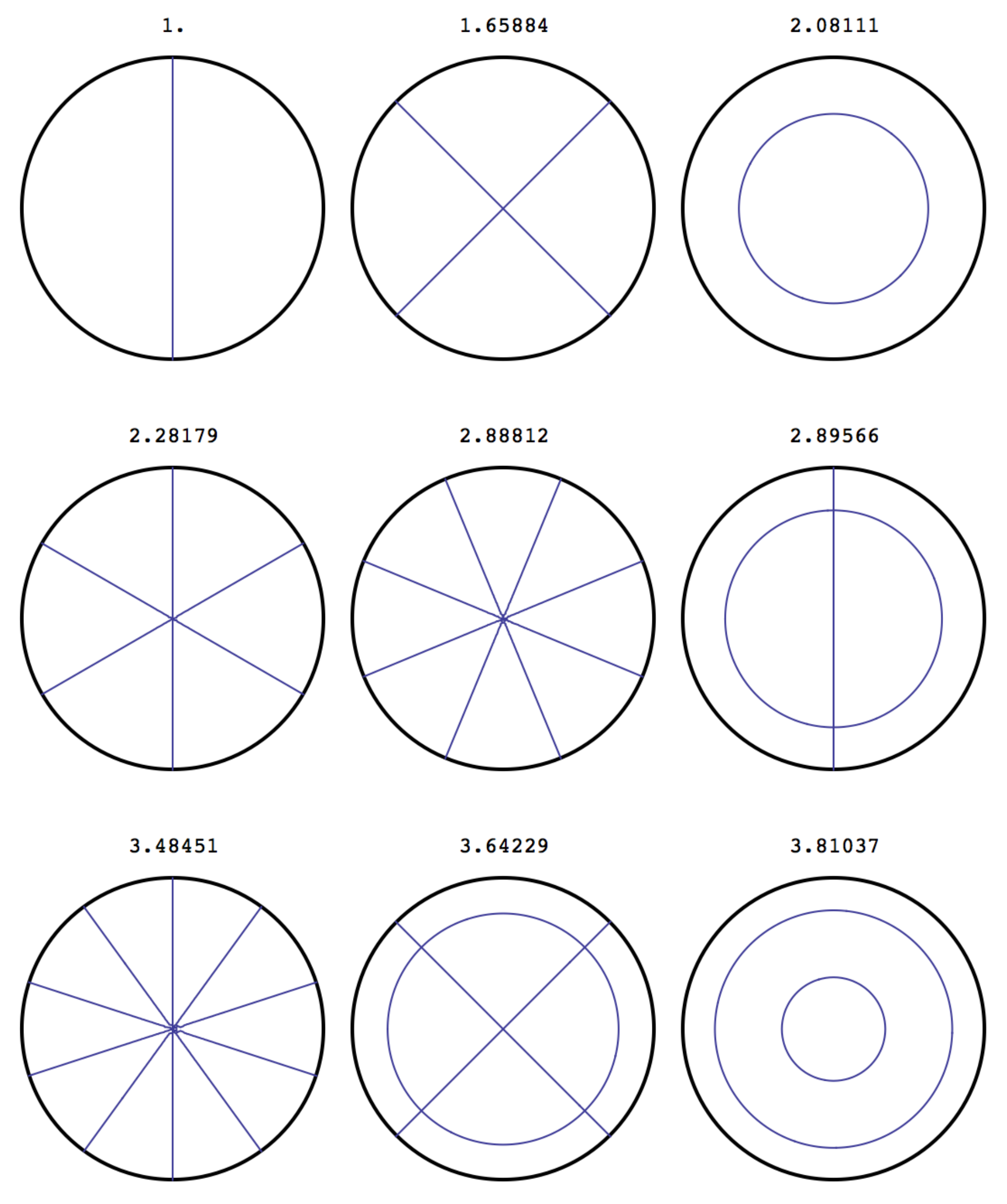}
\caption{Pressure node lines of the lowest squat cylinder modes; the number above is the frequency in units of the lowest mode frequency}
\end{figure}

The internal diameter of the actual pot was measured.  The resulting squat cylinder frequency values are indicated by the black vertical lines at the bottom of FIG.~15.  (That calculation won't be perfect because it ignores the presence of the speaker, microphone, tone ring, and other hardware inside.)  The additional grey lines, first appearing around 2300 Hz, are the calculated frequencies of the additional modes that involve wave components in the squat, perpendicular direction.

The inputs into the calculation are: the pot inner diameter and height, the compressibility of air (as encoded in the speed of sound), and Newton's Laws.  The success is a triumph of physics, but it is very old and well-known physics.  The ``$3/8''$ gap" purple line is the spectrum we want to understand because it represents an essential feature of the internal resonator banjo.  Of course, it is not the final sound of the banjo because those resonances have to couple to the head; the net effects on the head and the actual sound are subtle but certainly within the range of things people can distinguish.

The key to conceptual progress came with abandoning the attempt to relate the two curves in FIG.~15 by building up the height of the wall.  Rather, an enlightening starting point is a wall that leaves no gap between itself and the head.  Then the smaller inner cylinder and the outer annular region are distinct.  A small acoustical coupling between the two was introduced in the form of a $1/2'' \times 1/2''$ hole in the internal wall.  The result (and more) is in FIG.~17.

\begin{figure}[h!]
\includegraphics[width=6.5in]{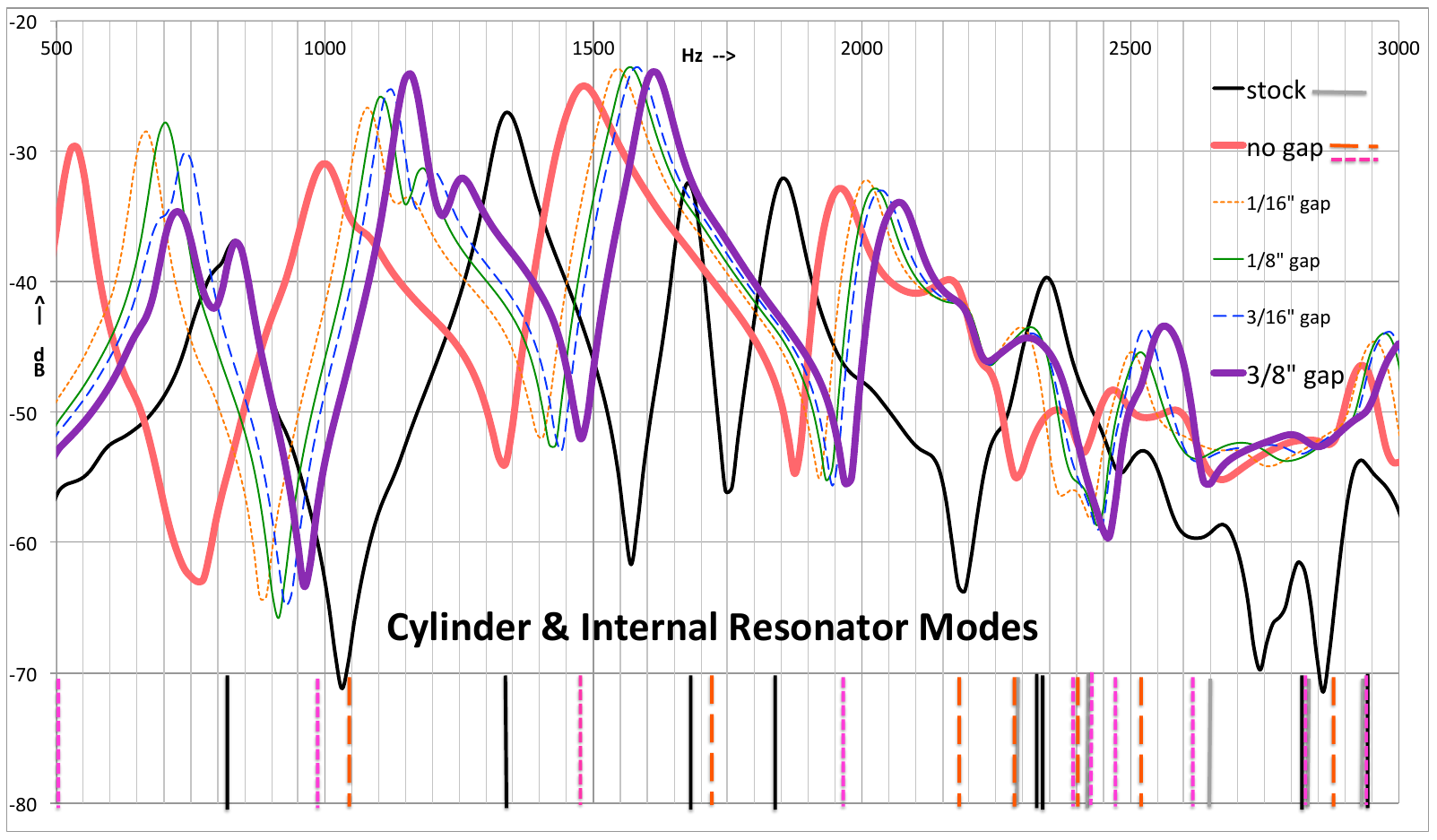}
\caption{Comparison of the stock pot \& the internal resonator with various gaps; lines at the bottom denote frequencies calculated from the actual physical dimensions}
\end{figure}

As in FIG.~14, the black curve comes from the stock pot scan, and the purple curve is for the $3/8''$ gap, i.e., the normal dimension internal resonator with a 2 $1/4''$ high wall.  The red curve is for the ``no gap" wall that seals up against the head.

I indicate on the bottom the calculated resonant frequencies for the ``no gap" system, under the assumption that the coupling of its two parts is weak enough to ignore.  The long dash orange/red lines are the standard cylinder mode frequencies, higher than the black ones simply by the ratio of the stock diameter to the internal resonator diameter, at least for the squat, lower frequency range.  The two cylinders have the same height, and the higher frequency contributions from waves in the perpendicular direction are added in also.

The calculated mode frequencies of the annular volume are indicated with short dash pink/red and use the approximation described in section II.  They begin around 500 Hz, which is substantially lower than the lowest mode of the stock cylinder.  Note that the smallest dimension of the annulus is $1.01''$ in the radial radial direction.  That is only first excited around 6700 Hz.

FIG.~17 also displays the measured spectra for intermediate values of the rim-wall-head gap, illustrating how the spectrum evolves continuously from no gap to its final $3/8''$ value.

In contrast, the intermediate spectra resulting from successively lower internal walls, i.e., going from the standard 2 $1/4''$ height down to zero, which is equivalent to the stock pot, are more confusing than enlightening.  (Remember that the back is sealed with solid plywood; there is no ``partial back" in this part of the analysis.)

\begin{figure}[h!]
\includegraphics[width=6.0in]{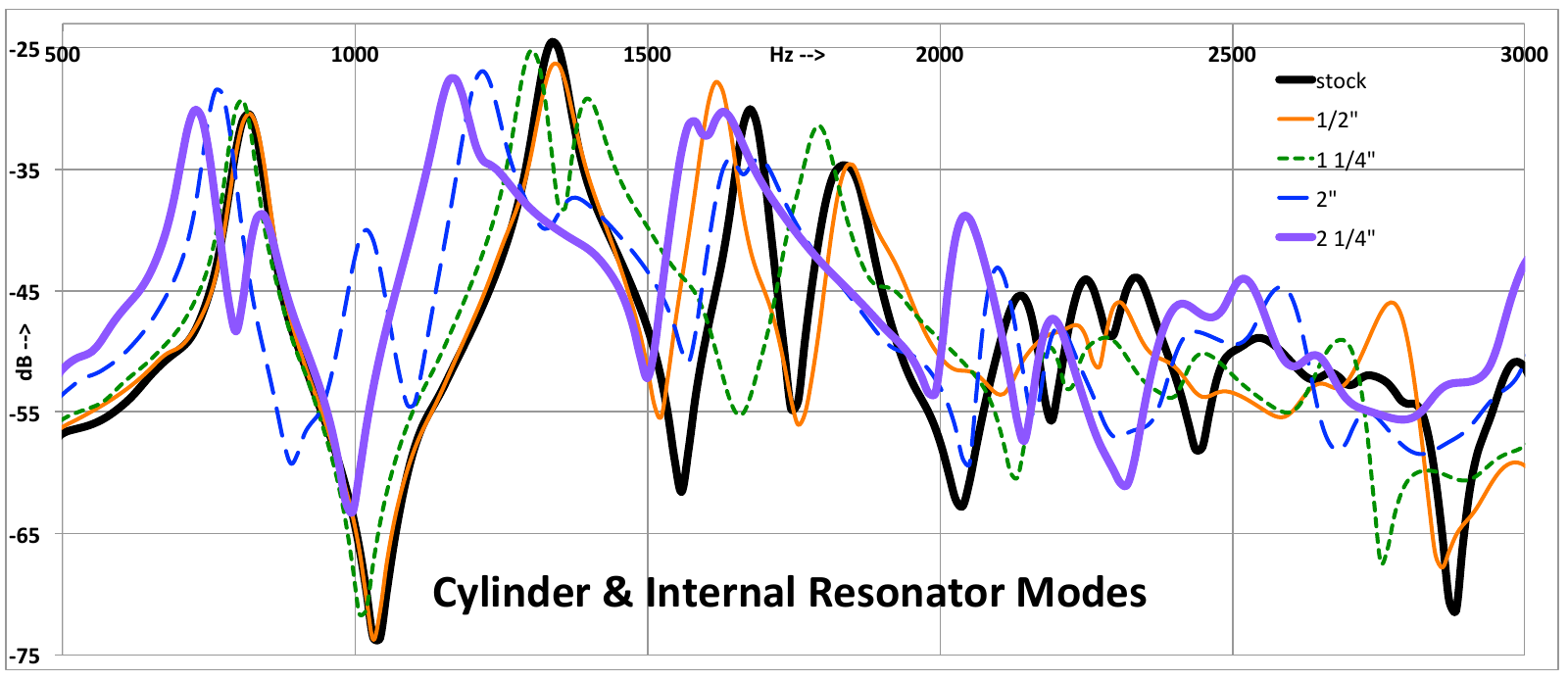}
\caption{Spectra for different values of the internal cylinder wall height}
\end{figure}

\noindent
These are the spectra displayed in FIG.~18.  The black and purple curves do not quite match those of FIG.~17.  The origin of this discrepancy is the following.  Between taking the data presented in each of the figures, the speaker and microphone were removed and re-mounted, without particular care to reproduce the original positions. The resonant frequencies of the pots  should be unchanged, but their individual strengths can differ significantly, particularly at the higher frequencies.  The point is that the speaker subtends an angle of about $13^{\text{o}}$ (in azimuth), and the microphone about $10^{\text{o}}$.  Their separation was about $26^{\text{o}}$.  These numbers can have a strong effect on how much a particular mode is excited and how well it is detected.

\section{Fully assembled banjo plucked and played}

The following are a few comparisons of two fully assembled banjos: a totally normal Goodtime and a fully Bacon-like modified Goodtime, i.e., with tone ring and internal 2 $1/4''$ resonator.  The strings, heads, and head tensions (as measured by a DrumDial) were the same.

Here is a sound recording and spectrograph of four typical single string plucks, with the other stings left free to vibrate.  The microphone was at $20''$ in front of the head.  All plucks were at the second fret.  The first one is the $4^{\text{th}}$ string of the normal Goodtime; the second is the  $4^{\text{th}}$ string of the Bacon-modified Goodtime; the third is the $1^{\text{st}}$ string of the normal Goodtime; and  the fourth is the $1^{\text{st}}$ string of the Bacon-modified Goodtime.  (If your reading is Web-enabled, the following links might be live; otherwise they should be retrievable.)  This is the sound file:

\centerline{\href{http://www.its.caltech.edu/~politzer/bacon/4-plucks.mp3}{http://www.its.caltech.edu/\url{~}politzer/bacon/4-plucks.mp3}}

\noindent and FIG.~19 is a spectrograph.
\begin{figure}[h!]
\includegraphics[width=6.5in]{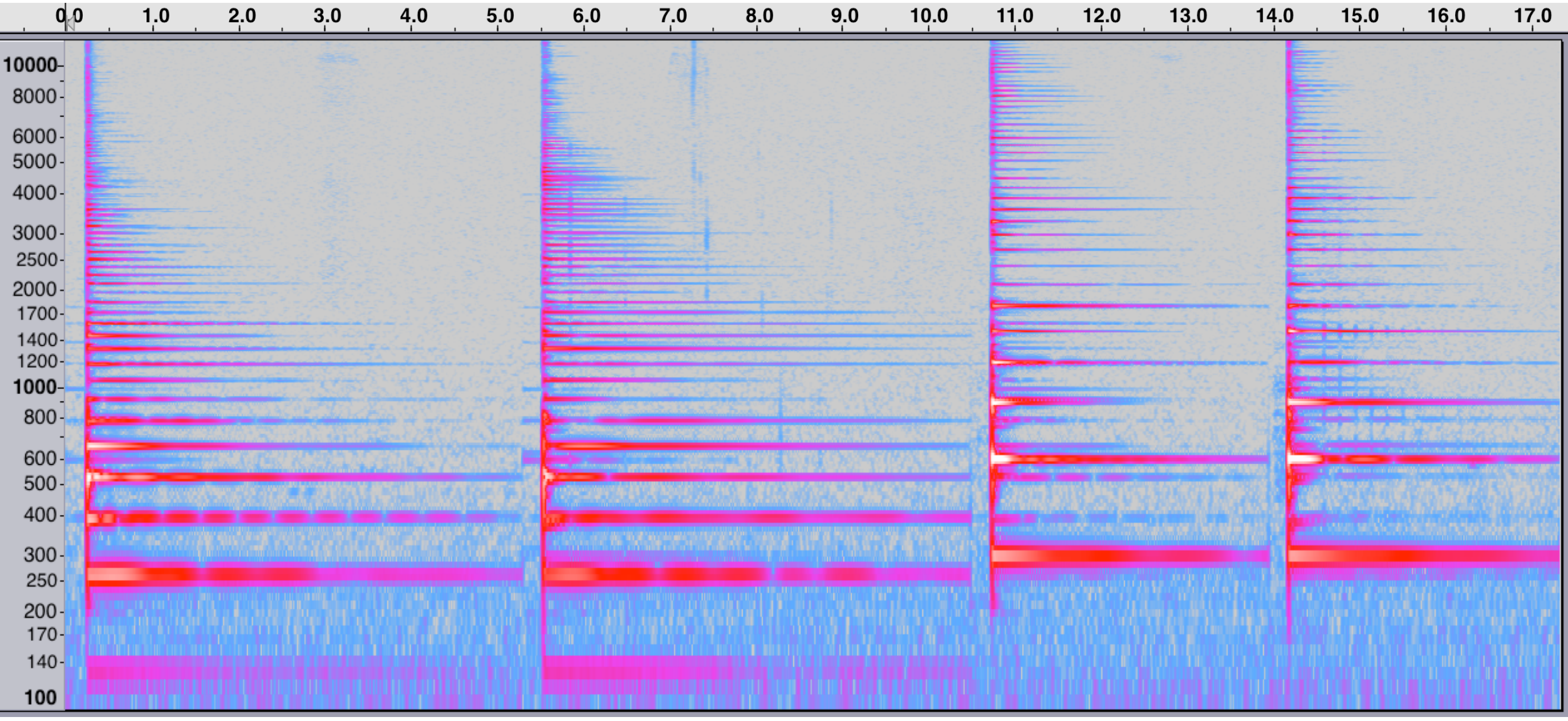}
\caption{Spectrograph of the four plucks described and linked above}
\end{figure}

Before commenting on the differences, I offer here brief samples of frailing on the two banjos:

\href{http://www.its.caltech.edu/~politzer/bacon/sample-A.mp3}{http://www.its.caltech.edu/\url{~}politzer/bacon/sample-A.mp3}

\href{http://www.its.caltech.edu/~politzer/bacon/sample-B.mp3}{http://www.its.caltech.edu/\url{~}politzer/bacon/sample-B.mp3}

Which is which is revealed here: \cite{AB} .

A frequency spectrum analysis for each entire 35 second played selection is displayed in FIG.~20.

\begin{figure}[h!]
\includegraphics[width=6.5in]{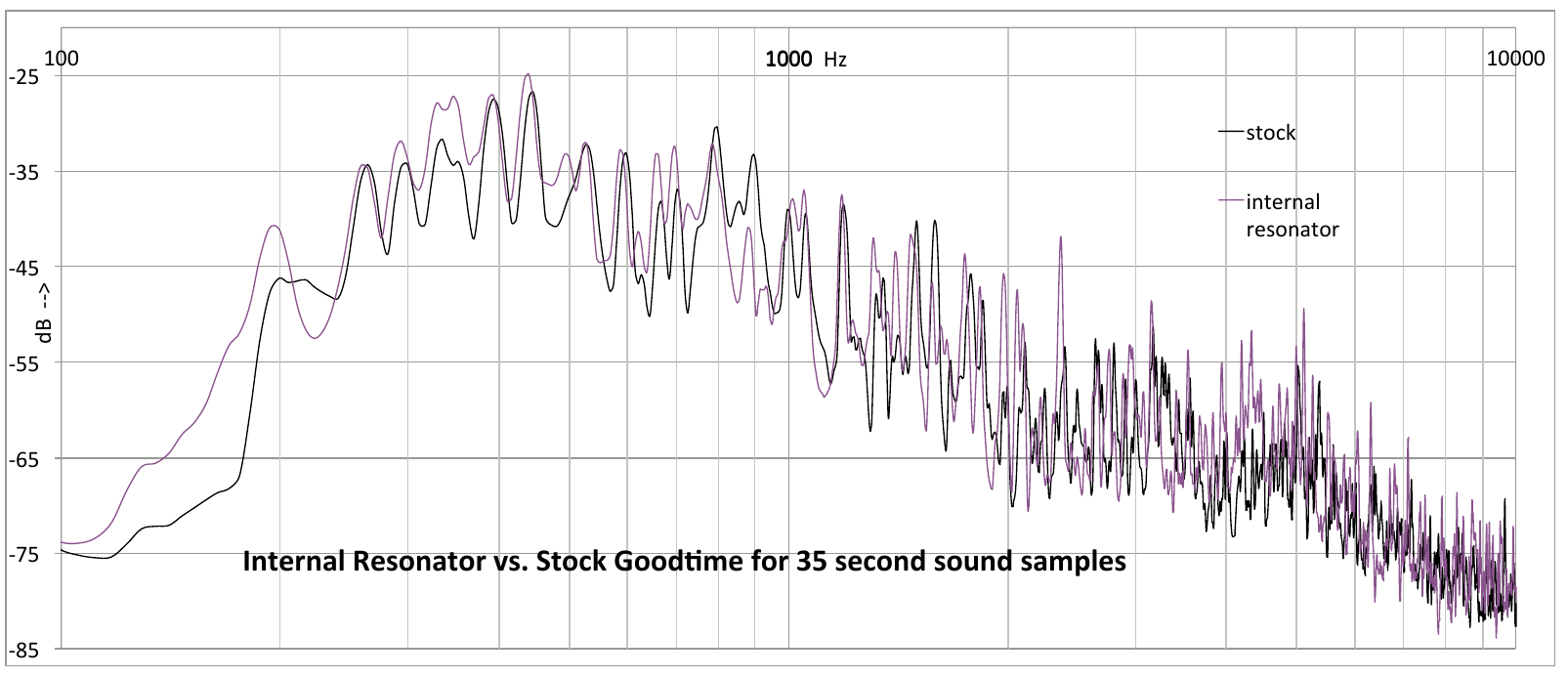}
\caption{Spectra of 35 seconds of frailing on a Goodtime stock {\it vs} internal-resonator-fitted banjo }
\end{figure}

The most obvious difference apparent in the FIG.~20 spectrum analysis is in the power below 200 Hz.  In that range, neither is very loud in an absolute sense.  But the extra strength provided by the Bacon design likely accounts for why some people mention that the sound of an internal resonator banjo is reminiscent of a $12''$ rim, rather than an $11''$ (which is its actual size).

The spectrograph helps to identify differences that can be heard in the plucks and in  the actual played sound samples.  Interestingly but not surprisingly, there's a lot going on.  And, gratifyingly, much can be traced to the particular physical distinctions identified in this and the accompanying tone ring study.

For the low open $4^{\text{th}}$ string, the Bacon modifications make the fundamental stronger but the next six harmonics weaker.  Then, above 1200 Hz, the Bacon-style banjo has more power {\it and} sustain.  On the open $1^{\text{st}}$ string, the Bacon is mostly comparable or even a bit stronger in power and sustain until around 1700 Hz, after which it's the other way around.

\section{Conclusion}

Apparently, the partial back and the annular region provide lower body-air resonances than any that are present on the stock, open-back banjo.  And the combined central cylinder and annulus provide a richer spectrum of coupled resonances than the simpler pot.  Hearing the effect of the lower resonances is straightforward.  

The richer spectrum aspect is more subtle.  Certainly, more resonances, closer together, give a more even response as a function of driving frequency.  But the representation of a system by its spectrum loses all reference to time development.  And, for plucked (or struck) strings, the time evolution is an essential characteristic.  Every aspect of the sound is transient.  However, the transients of coupled oscillating systems are relatively neglected in physics and engineering education and are not very widely understood or appreciated.  Having resonances that are nearby in frequency but which arise in different parts of a system allow a variety of interesting phenomena.\cite{coupled-damped}  The actual resonant frequencies can depend on the coupling strength. Energy can flow from one to the other and back.  This can extend the lifetime or sustain.  It can give rise to beats.  So, the internal resonator design not only reduces some of the banjo's most shrill bark, it increases at least some of the sustain and presents a richer, more complex sound.

\bigskip
\bigskip

\bigskip

\bigskip

\centerline{\bf NOTES}

%\bigskip

\bigskip


\begin{thebibliography}{99}

\bibitem{patent}Fred Bacon's internal resonator patent: \href{http://www.google.com/patents/US823985}{http://www.google.com/patents/US823985}

\bibitem{rayleigh}John William Strutt, $3^{\text{rd}}$ Baron Rayleigh, {\it The Theory of Sound}, 1887 \& 1896, still in print  and widely available in various formats.  It's not particularly easy to read or use as a reference, but it is dense with insights and one of the most inspiring physics books I've ever encountered.

\bibitem{periodic-boundary}``Identified ends" means that whatever goes out one end goes in the other --- and {\it vice versa}.  For example, the lowest frequency standing waves go around the long way, with wavelength equal to the circumference.  The two node positions are not determined.

\bibitem{danish}The Danish Bottle Boys are worth a listen.  My on-line favorite of theirs is \href{https://www.youtube.com/watch?v=NkbZlautuUc}{https://www.youtube.com/watch?v=NkbZlautuUc} .

\bibitem{openback}D.~Politzer, {\it The Open Back of the Open-Back Banjo}, HDP: 13 -- 02, \newline www.its.caltech.edu/\url{~}politzer

\bibitem{tone-ring}D.~Politzer, {\it A Bacon Tone Ring on an Open-Back Banjo}, HDP: 16 -- 01, www.its.caltech.edu/\url{~}politzer

\bibitem{Goodtime}I originally chose Deering Goodtime banjos for my acoustics investigations because 1) they are about as identical as wood objects can be, being a combination of CNC fabrication and high quality hand finishing; 2) they are quality instruments; and 3) they are relatively inexpensive.  When I fist approached Greg Deering, requesting some special items and perhaps a deal on the price, he immediately offered to provide me with whatever I needed.  He has been of great help ever since, including advice and fabrication related to this project.

\bibitem{cylinder-modes}e.g., M.~J.~Moloney, {\it Plastic CD containers as cylindrical acoustical resonators,} Am.~J.~Phys. {\bf 77} (10) 882 (2009); DOI: 10.1119/1.3157150.

\bibitem{coupled-damped}D.~Politzer,{\it The plucked string: an example of non-normal dynamics}, HDP: 14 -- 04, {\it Am. J. Phys.} {\bf 83} 395 (2015), doi:10.1119/1.4902310 or www.its.caltech.edu/\url{~}politzer; {\it Zany strings and finicky banjo bridges}, HDP: 14 -- 05, www.its.caltech.edu/\url{~}politzer

\bibitem{AB}B is the stock Goodtime, and A is a Goodtime fitted with a bacon-style tone ring and internal resonator.


\end{thebibliography}
\end{document}